# ANALYSIS OF PAIN HEMODYNAMIC RESPONSE USING NEAR-INFRARED SPECTROSCOPY (NIRS)

Raul Fernandez Rojas[1], Xu Huang[1], Keng Liang Ou[2], Dat Tran[1] and Sheikh Md. Rabiul Islam[1]

[1]Faculty of Education, Science, Technology and Mathematics, University of Canberra, Australia
[2]College of Oral Medicine, Taipei Medical University, Taiwan

***ABSTRACT***

*Despite recent advances in brain research, understanding the various signals for pain and pain intensities in the brain cortex is still a complex task due to temporal and spatial variations of brain haemodynamics. In this paper we have investigated pain based on cerebral hemodynamics via near-infrared spectroscopy (NIRS). This study presents a pain stimulation experiment that uses three acupuncture manipulation techniques to safely induce pain in healthy subjects. Acupuncture pain response was presented and Haemodynamic pain signal analysis showed the presence of dominant channels and their relationship among surrounding channels, which contribute the further pain research area.*

## 1. INTRODUCTION

Pain is a subjective experience and every individual experiences it in different ways. Pain perception can be range from emotional pain such as the loss of a love one to physical pain such as a broken leg. In addition, pain sensation as the product of dedicated neural mechanisms continues to be a topic of debate. But what is pain? It seems to be very hard to define. According to Perl [1] "With the benefit of the past two centuries of scientific work and thought, can one define pain? Considering the evidence, it seems reasonable to propose pain to be both a specific sensation and an emotion, initiated by activity in particular peripheral and central neurons. Pain shares features with other sensations, but the strong association with disposition is special."

There is a number of publications about pain research [2-4] to answer questions such as: how to interpret it, how to control it, or how to use it to indicate human body health status. In this paper we focus on the instance that an unconscious patient under medical examination cannot express level of pain experience in that moment. For this reason, this research tries to find a relationship between the pain intensity and the hemodynamic response to understand the patient's feelings. By finding this "silent communication" between pain and brain hemodynamics, medical services will be able to provide a better solution for such pain sensation.

Near infrared spectroscopy (NIRS) is employed in our experiments. NIRS can offer simultaneous measurements from dynamic changes of Oxyhemoglobin (HbO) and Deoxyhemoglobin (HbR) in the brain cortex with a reasonable temporal resolution (less than one second) and spatial resolution (less than 3 cm), while Total haemoglobin (HbT) is obtained by the difference between HbO and HbR. Our experimental study makes use of changes in HbO to represent neural activity in the human brain. In addition, acupuncture is also used to induce pain to patients in a safe manner.

This paper is organised as follows. Section 2 gives a review of the equipment and experimental setup. The related analysis methods will be discussed responding to the experimental setup. In section 3, the experimental results will be investigated, including dynamic NIRS response to acupuncture stimulation, pain responses of HbO2 and HbR to acupuncture stimulation, pain responses to different acupuncture manipulations, dominant pattern analysis, and direction of pain information. In section 4, conclusions are presented.

## 2. Materials and Methods

### 2.1. Equipment

Hemodynamic measurement was performed using an optical topography system Hitachi ETG-4000 (Hitachi Medical Corporation). The topography system uses near-infrared spectroscopy (NIRS) to investigate cerebral hemodynamics. Optical topography makes use the different absorption spectra of oxygenated and deoxygenated haemoglobin in the near infrared region. The NIRS system produces 695 and 830 nm NIR signals through frequency-modulated laser diodes. These NIR signals are transmitted to the brain using optical fibre emitters (Figure 1, red circles). Near-infrared light penetrates head tissue and bone, generally it reaches 2 to 3 cm into the cerebral cortex. Once NIR light reaches the cerebral cortex, it is absorbed by haemoglobin, while the non-absorbed NIR signals are reflected to the source, where it is sampled by a high-sensitivity photodiode (Figure 1, blue circles). NIR light between emitters and detectors is sampled at a given time point named channels (Figure 1, numbered squares). Since Oxyhemoglobin (HbO) and Deoxyhemoglobin (HbR) absorb NIR light differently, two wavelengths of light (695 and 830 nm) are used. In this way, it is possible to read these two types of haemoglobin simultaneously. The sample frequency used in this experiment was of 10Hz (10 samples per second). The configuration for this experiment was using two probes of 12 channels to measure neurologic activity of the head (parietal/temporal region). Figure 1 shows the 24 channel configuration used in the study, channels 1 to 12 represent the right hemisphere, while channels 13 to 24 represent left hemisphere. Signals in between channels were computed from the surrounding channels.

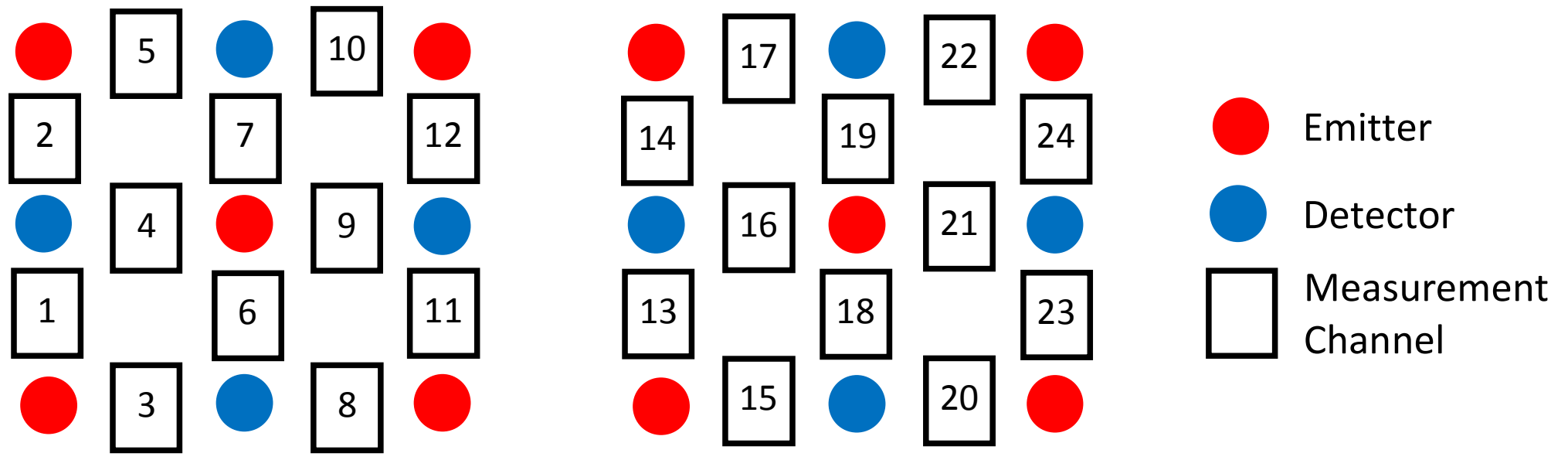

**Figure 1.** Channel configuration, right hemisphere (channels 1-12) and left hemisphere (channels 13-24).

### 2.2. Experiment Setup and Subjects

The experiment was designed by the School of Oral Medicine of Taipei Medical University (TMU) in collaboration with the University Of Canberra (UC). In the present study, six healthy individuals (2 females, 4 males) participated in the experiment, aged 25 to 35 years old. All participants provided written consent and the experiment was approved by the Ethics Committee of TMU. The experiment was carried out in the Brain Research Laboratory at TMU in a quiet, temperature (22-24°C) and humidity (40-50%) controlled room. The experiments were done in the morning (10:00am-12:00pm) and each experiment lasted around 30 minutes. Quantitative data was collected using the ETG-4000 with the patients sat down in an ergonomic chair near the topography system.

The experiment was designed to recognise pain stimulation and pain release in patients through hemodynamic responses. In order to identify the pain through NIRS, acupuncture was used to induce pain stimulation in a safe manner. Brand new acupuncture needles were used for each experiment, and using traditional Chinese acupuncture techniques that were performed by an acupuncturist of TMU Hospital. The puncture point used for stimulation was the “Hegu Point”, located on top of the hand, between the thumb and forefinger. This acupuncture point (acupoint) is known by its property to relieve pain, especially headaches and toothaches. This acupoint is also used to reduce fever, eliminate congestion in the nose, stop spasms, and decrease toothache. A western name for this acupoint is “the dentist’s point” because it can stop tooth pain while moist the throat and tongue. This point was used because it is an area of easy access and the hand can be set aside while the patient is relaxed on the chair. Each patient was punctured on both hands (Figure 2), one hand on a day and the opposite hand on another day; each hand was treated as a separated experiment. The data base was organised of 12 data sets of changes in Haemoglobin files, two sets (right and left hands) per each subject.

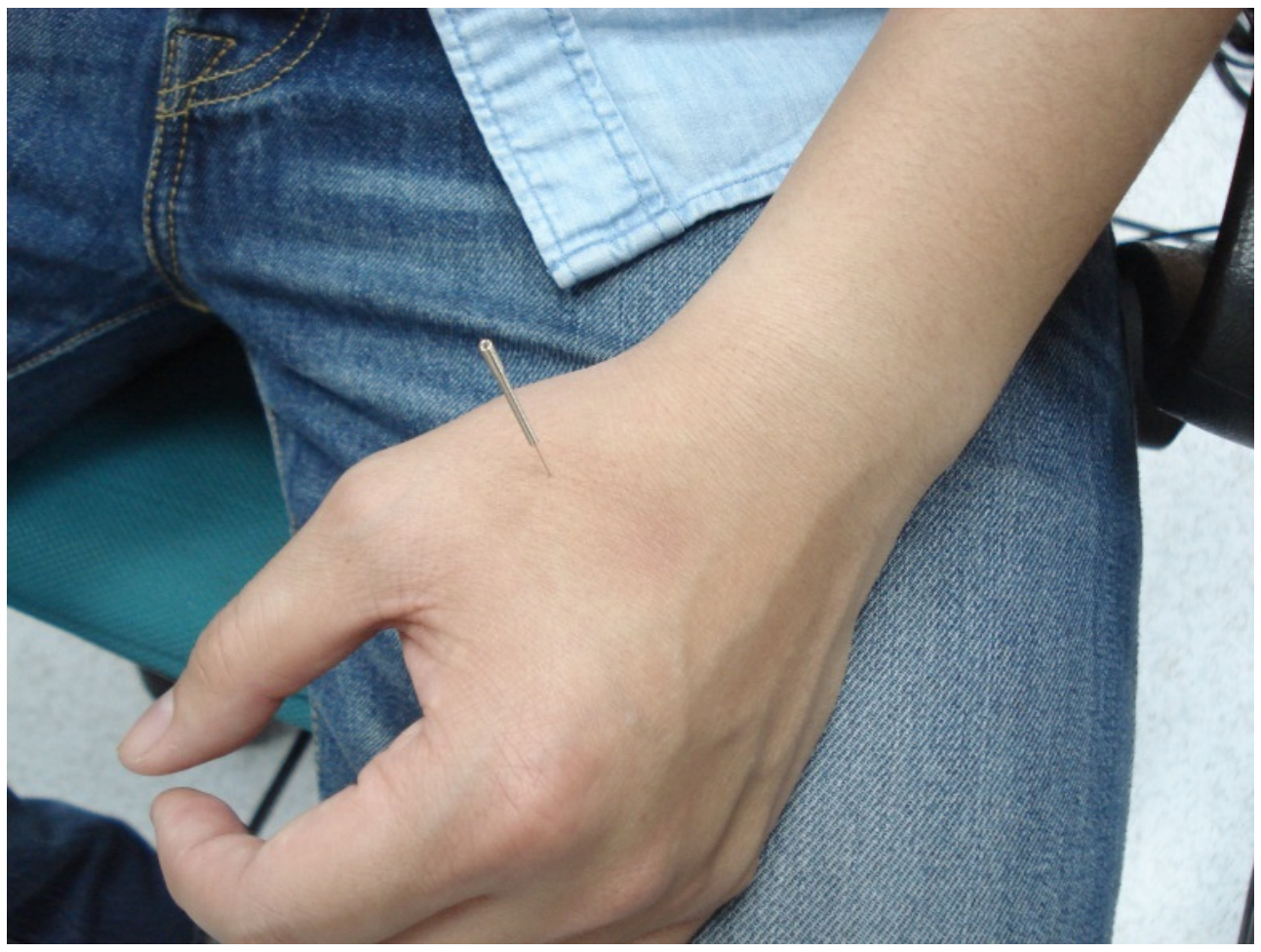

Figure 2. Patient with acupuncture needle in Hegu point.

Three types of pain stimulations (acupuncture techniques) were applied in the experiment, needle insertion (Task 1, T1), needle twirl (Task 2, T2), and needle removal (Task 3, T3). The first type of stimulation (T1) was carried out 30 seconds after the start of experiment and it lasted for 6 seconds. The second type of stimulation (T2) was applied for 30 seconds (rest time) after T1 and it lasted for 10 seconds, this stimulation was repeated three times after 30 second resting time. The last stimulation (T3), was carried out after the third application of task 2, and it lasted for 5 seconds with a 30 seconds recovery time and 15 seconds post-time to finish the experiment.

The patients were explained about the acupuncture procedure and experiment. In addition, patients were given a brief explanation of side effects of acupuncture in case they had any symptoms during the experiment; no side effects were reported during and after experiment. After the briefing, the patient was told to sit down on the chair, told to relax and close the eyes to reduce visual evoked stimulation. The optical topography system was set to record after all conditions were met and patient was relaxed.

## 2.3. Stages of Experiment

Figure 3 exhibits the three types of acupuncture stimulations applied in the experiment. The image shows the changes of HbO, HbR, and (HbT) concentration during the different stages of the experiment; for illustration purposes data was taken from subject 4 in channel 19. The first stage of the study is the pre-time which is used to adapt the patient to the experiment and obtain baseline for the haemoglobin signals, the pre-time lasted for 30 seconds. The second stage (T1, task 1) is started immediately after pre-time stage and it begins the first pain stimulus (needle insertion), and it is done within 6 seconds. After the pain stimulation, patient is given a resting time of 30 seconds (Rt1), which allows the haemoglobin concentration to return to baseline for the next stimulation. The next task (T2) is needle twirl, it lasted 10 seconds and is followed by a 30 second resting time (Rt2); this task was repeated two more times including resting time (Rt3, Rt4) of 30 seconds each. Task three (T3) is started immediately after Rt4 and it lasted for 5 seconds; in this task the needle is removed from the hand. The last recovery time (Rt5) allows the patient to recover from the previous pain stimulus and also haemoglobin signals return to baseline, it lasts for 30 seconds. The last stage in the experiment is the Post-time and lasts for 15 seconds, after the end of this time the equipment stops recording data.

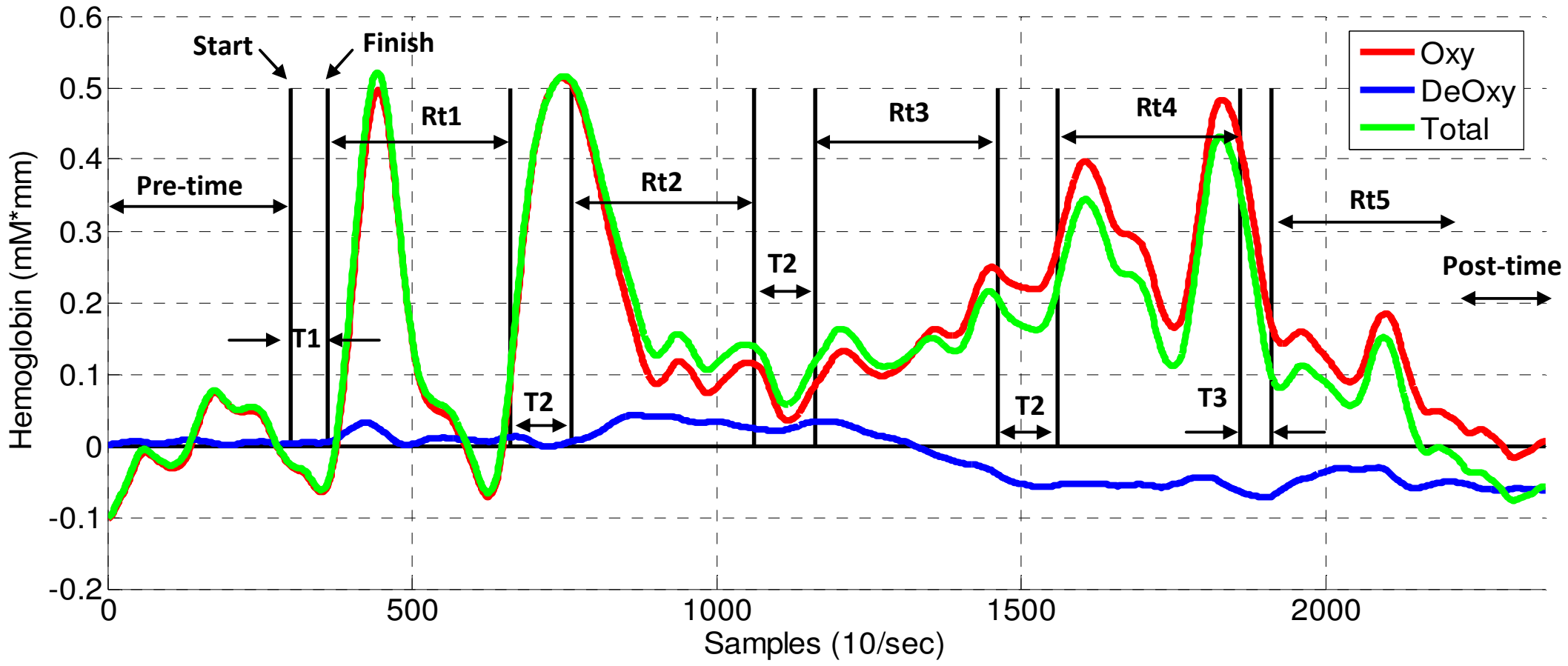

Figure 3. Description of tasks of the experiment, Rt=rest time, T=task.

## 2.4. NIRS colour representation

Figure 4 shows a colour representation of a typical response from a sample of HbO intensity with the 24 channels of subject number 4. The colour representation helped to distinguish pattern trends and focus on areas where the dynamic response was more active. Left figure represents the channel intensities of the right hemisphere, while the right figure represents the channel intensities of the left hemisphere. The colour intensity runs from -0.3 (blue/cold) to +0.5 (red/hot); being the blue colour the lower measurements of HbO, while the red colour represents the higher values of HbO. For visual analysis, values above +0.5 and below -0.3 were represented within this range.

## 2.5. Data Analysis

Signal Analysis was performed by two signal processing techniques. The first method was the cross correlation analysis of raw data to identify time-related channel similarities. While the second method was the Horn and Schunk optical flow (OF) algorithm, this method was used to identify patterns within both hemispheres. Both methods are described as follows.

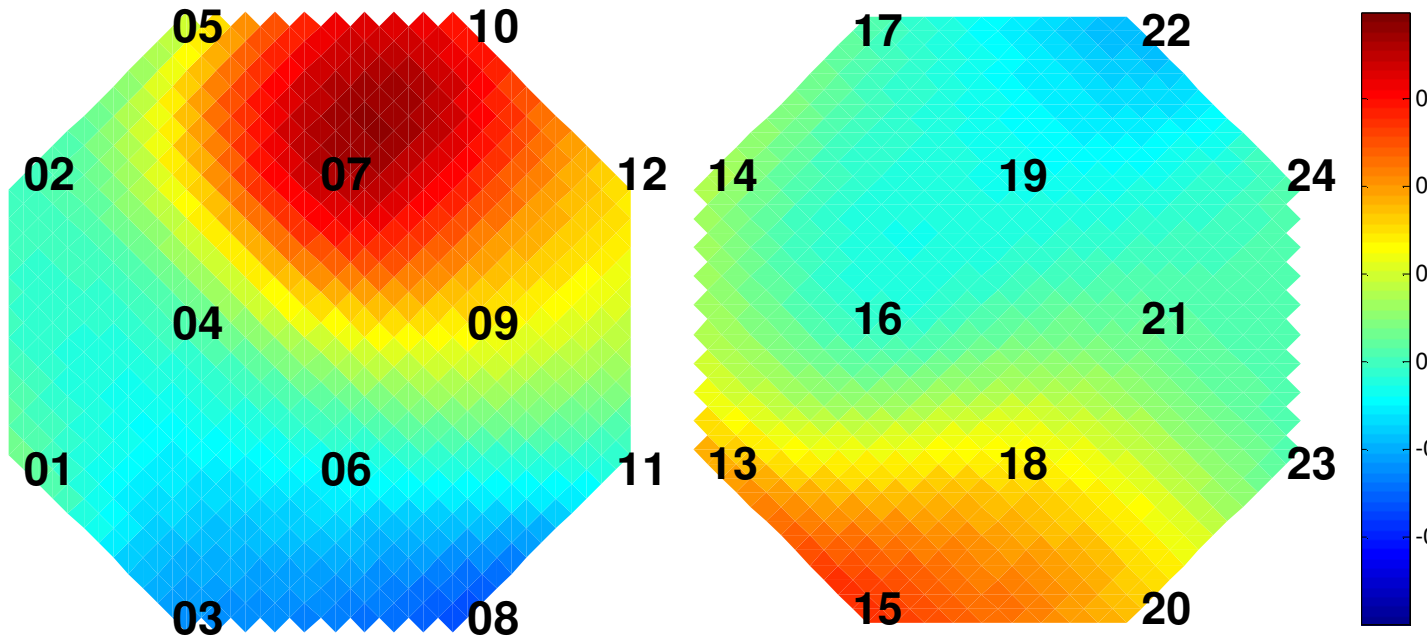

Figure 4. Channel intensity, right (channels 1-12) and left (channels 13-24) hemispheres.

### 2.5.1. Cross Correlation Analysis

Data analysis was first performed by comparing channels on both sides of the head. Cross correlation function was computed between channels 1-12 in right probe and 13-24 in left probe. The cross correlation between two input signals (channels) is a kind of template matching. This measure of temporal similarity of two signals was done by computing a time-shifting along one of the input signals. Equation 1 defines the cross correlation between two waveforms *x(t)* and *y(t)* as follows:

$$r_{xy}(\tau) = \sum_{-\infty}^{\infty} x(t)y(t-\tau) \qquad (1)$$

Where $\tau$ is the time-lag between *x(t)* and *y(t)*, the value of $r_{xy}$ denotes the difference (lag/lead) between channel signal *y(t)* and channel signal *x(t)*. The cross correlation value between two channels in the same probe is done from -40 sec to +40 sec at a rate of 10 samples per second.

### 2.5.2. Optic Flow Analysis

The optical flow is defined as the "flow" of pixel values at the image plane in time varying images [5]. Ideally the optical flow corresponds to the motion field, however in some situations the optical flow is not always equal to the motion field. To overcome any problems in the calculation of the optical flow, assumptions have to be defined. Firstly, it is assumed that the surface of the image being observed is flat to avoid variations in brightness due to shading effects. Secondly, the incident illumination is uniform across the surface. Thirdly, it is also assumed that the optical flow varies smoothly and has no spatial discontinuities.

We calculate optical flow from the apparent motion at the image plane based on visual perception with the dimensions of the image velocity of such as motion. We define the optical flow vector $\vec{v} = (u, v)$, where $u$ and $v$ represent the $x$ and $y$ components of the optical flow vector at a point on the image. As the optical flow is estimated from two consecutive images, it appears as a displacement vector $\vec{d} = (d_x, d_y)$ [6]. Where $d_x$ and $d_y$ denote the $x$ and $y$ components respectively of the displacement vector at a point.

The Horn and Schunk algorithm [5] implemented in this work is based on differential solving schemes. It computes image velocity from numerical evaluation of spatiotemporal derivatives of image intensities. Let denote image intensity (or brightness) at the point *(x,y)* in the image plane at time *t* by *E(x,y,t)*. Thus, the image domain is consequently assumed to be differentiable in space and time. The basic assumption in measuring image motion is that the brightness of a particular point in the time-varying image is constant, as a result,

$$E(x, y, t) = E(x + \delta x, y + \delta y, t + \delta t) = 0, \qquad (2)$$

where $\delta x$ and $\delta y$ are the displacement in *x* direction and *y* direction, respectively, after a short interval of time $\delta t$. This assumption brings to the following condition, known as the "Optical Flow Constraint Equation". If brightness varies smoothly with *x*, *y*, and *t*, Equation (2) can be expanded in a Taylor series and obtain

$$E(x, y, t) + \delta x \frac{\partial E}{\partial x} + \delta y \frac{\partial E}{\partial y} + \delta t \frac{\partial E}{\partial t} + e = E(x, y, t), \tag{3}$$

where *e* contains second and higher-order terms in $\delta x$, $\delta y$, and $\delta t$. Subtracting $E(x, y, t)$ on both sides, dividing by $\delta t$, and taking the limit as $\delta t \to 0$, we have

$$\frac{\partial E}{\partial x}\frac{dx}{dt} + \frac{\partial E}{\partial y}\frac{dy}{dt} + \frac{\partial E}{\partial t} = 0 \tag{4}$$

where $\frac{dx}{dt} = u$ and $\frac{dy}{dt} = v$, and using the abbreviations $E_x = \frac{\partial E}{\partial x}$, $E_y = \frac{\partial E}{\partial y}$, and $E_t = \frac{\partial E}{\partial t}$, we finally obtain Equation (5), which is called the Optical Flow Constraint Equation

$$E_x u + E_y v + E_t = 0. \tag{5}$$

The components u and v are used to approximate the velocity component in the direction of the spatial gradient of the image intensity. Therefore, this is an under-constrained equation, since only the motion component in the direction of the local gradient of the image intensity function may be estimated, this is known as "the aperture problem". It can be seen rewriting Equation (5) in the form,

$$(E_x, E_y) \cdot (u, v) = -E_t, \tag{6}$$

thus the component of the movement in the direction of the brightness gradient $(E_x, E_y)$ is

$$-\frac{E_t}{\sqrt{E_x^2 + E_y^2}}. \tag{7}$$

From Equation (7), it is seen that we cannot determine the component of the optical flow at right angles to the brightness gradient. As a result, the flow velocity *(u,v)* cannot be computed locally without introducing additional constraints [5].

Horn and Schunck introduced a smoothness constraint, supposing that the motion field is smooth over the entire image domain [5]. This constraint, measures the departure from smoothness in the velocity flow. To express the smoothness constraint is required to minimize the square of the magnitude of the gradient of the optical flow velocity as follows,

$$e_s = \iint ((u_x^2 + u_y^2) + (v_x^2 + v_y^2))dxdy \tag{8}$$

the error in the optical flow constraint equation can be expressed as,

$$e_e = \iint (E_x u + E_y v + E_t)^2 dxdy \tag{9}$$

Horn and Schunck's algorithm computes an estimation of the velocity field *(u, v)* that minimizes the sum of the errors (Equation 9) for the rate of change of image brightness in equation (3), and the measure of the departure from smoothness in the velocity flow, equation (8) [5]. By combining equation (3) and (8) with a suitable weighting factor, and by using calculus of variation to minimize the total error, it is formulated an iterative solution for optical flow:

$$u^{n+1} = \bar{u}^n - \frac{E_x\left(E_x \bar{u}^n + E_y \bar{v}^n + E_t\right)}{\propto^2 + E_x^2 + E_y^2} \tag{10}$$

$$v^{n+1} = \bar{v}^{n} - \frac{E_y\left(E_x\bar{u}^{n} + E_y\bar{v}^{n} + E_t\right)}{\propto^2 + E_x^2 + E_y^2} \qquad (11)$$

Where superscripts indicate the iteration number, subscripts refer to derivation, and α is a positive constant know as smoothness factor.

# 3. Experimental Results

## 3.1. Dynamic NIRS response to acupuncture stimulation

The hemodynamic response was visualised by computing the colour representation in both probes. Ten serial samples from subject 4 based on HbO data from the 24 channels are presented in two separated images. Left image represents channels 13 to 24, while right image represents channels 1 to 12; the ten samples were taken every 22 seconds. Figure 5B shows the hemodynamic response in channel 1 while Figure 5C shows the hemodynamic response in channel 17. Red concentrations in Figure 5A indicate higher hemodynamic response to acupuncture stimulation. The increase and decrease of HbR was less significant than the response of HbO2, this effect was observed on both sides of the head. In addition, all patients showed a similar response pattern however different lags and magnitudes among channels were obtained for each subject. Hemodynamic response to acupuncture stimulation is explained as follows.

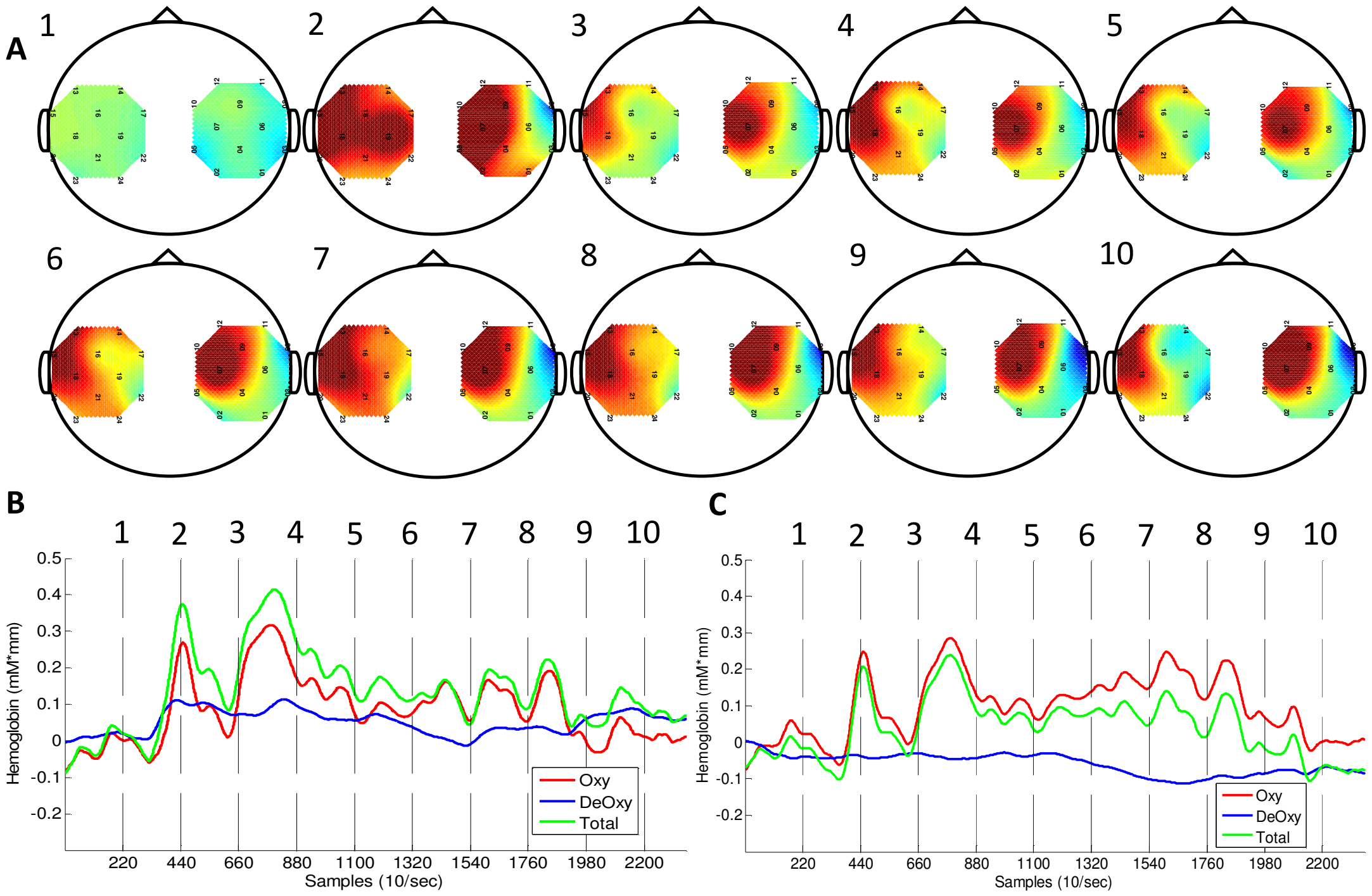

Figure 5. Dynamic haemoglobin response during experiment.

(A) Ten sequential NIRS images, obtained every 22 seconds, and numbered from 1 to 10. The images show the dynamic change of HbO trough 24 channels. The two images represent each side of the cerebral hemisphere.

(B) The dynamic trace recorded from channel 1 on the right side of the head. The number along upper Y axis corresponds to the numbered images in A.

(C) The dynamic trace recorded from channel 17 on the left side of the head. The number along upper Y axis corresponds to the numbered images in A.

Figure 6 exhibits the pain dynamic changes of haemoglobin (Hb) of Oxygenated Hb (HbO, in blue), Deoxygenated Hb (HbR, in red), and Total Hb (HbT, in green); For illustration purposes, data was taken from channel 3. Markers (dot lines) show the start and end of needle insertion stimulation (task 1), with a duration of 6 seconds. HbT is the combination of HbO and HbR, in other words is the sum of HbO and HbR. In a typical response after each acupuncture stimulation, the HbO presented a slight drop followed immediately for a constant rise to reach a peak, then a gradual decrease to resting level. On the other hand, HbR presented a slight increase then a moderate decrease to reach its lowest point; this variation was contrasting the reaction of HbO. In addition, HbT was dependent of both HbO and HbR, therefore it varied according to the behaviour of these two signals. It is to note that the estimations of HbO are more accurate than HbR and HbT [7, 8], for this reason HbO was used to present and analyse the data from this section onwards.

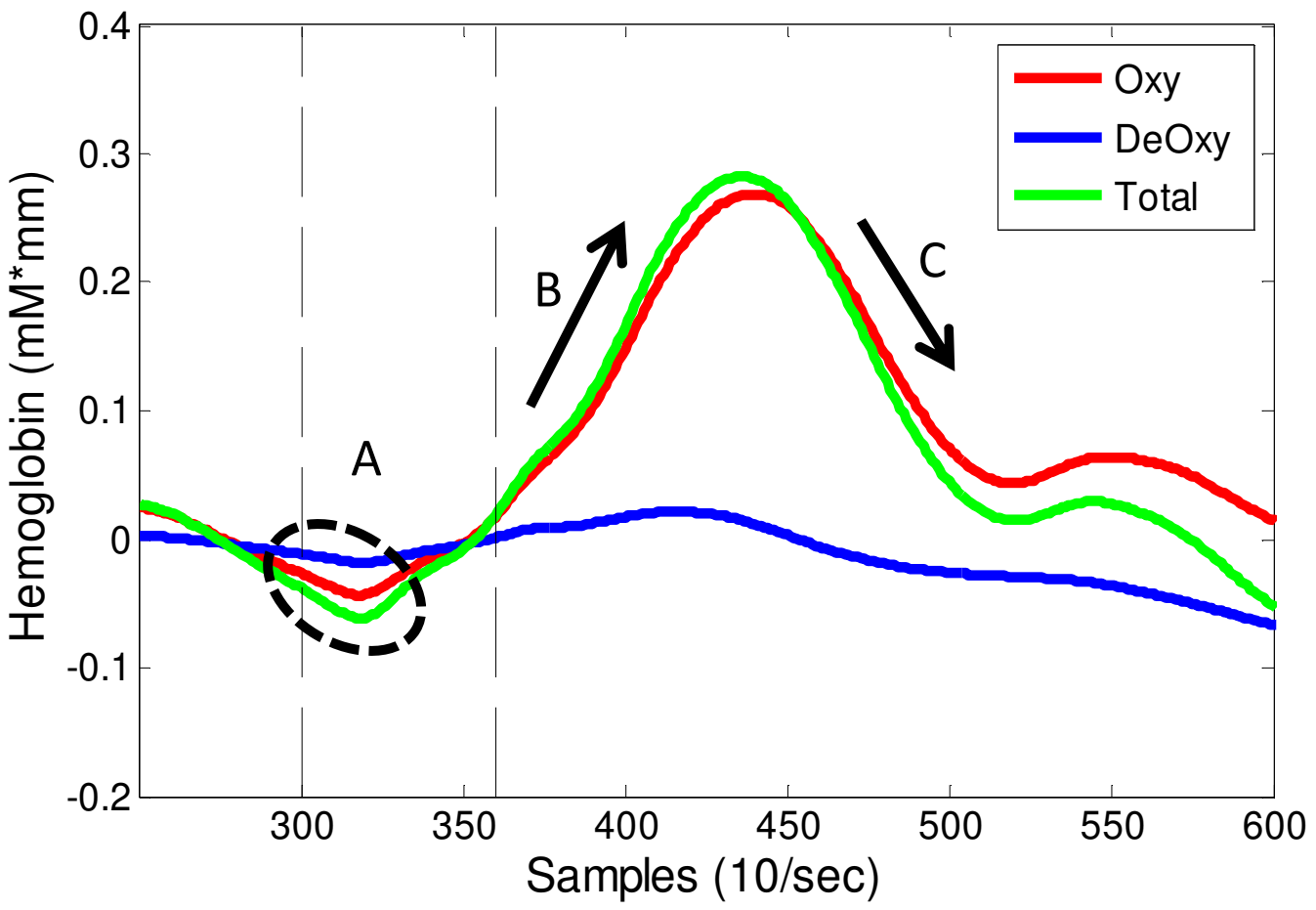

Figure 6. Distinctive dynamic response obtained after acupuncture stimulation. A) small drop of HbO after stimulation, B) followed by a constant increase until reach a peak, C) then a gradual decrease to resting level.

## 3.2. Pain responses of HbO to Acupuncture Stimulation

Needle insertion (task 1) was the first acupuncture manipulation in the experiment. It was applied at thirty seconds after commencement of experiment and lasted for six seconds. Each acupuncture manipulation was applied within the time indicated by the dot lines (markers) in Figure 7, four channels were used only for illustration purposes. Once the insertion was done (within samples 301 and 361), the needle was twirled until *de-qi* (Chinese for obtaining the qi or arriving at the qi) was reached. Once the patient manifested the numbness/heaviness in the arm, the acupuncturist stopped the stimulation. This *de-qi* sensation was approximately reached three seconds after needle insertion. Figure 7 shows the acupuncture pain response after needle insertion (T1). Immediately after reaching the highest pain response, approximately after 14 seconds, HbO dropped towards baseline. It is worth mentioning that in all subjects the highest peak was exhibited after this first stimulus (needle insertion). This drop-rise-drop behaviour was exhibited on both sides of the head (Figure 5B,C). Needle insertion was followed by a resting time of 30 seconds. This time was used for the patient to rest after the stimulus and let the haemoglobin signals go back to base line. It is worth mentioning that there is delay in reaction time between the physical stimulation and the NIRS signal among all the stimulations, and this is consistent with previous research using NIRS [9, 10].

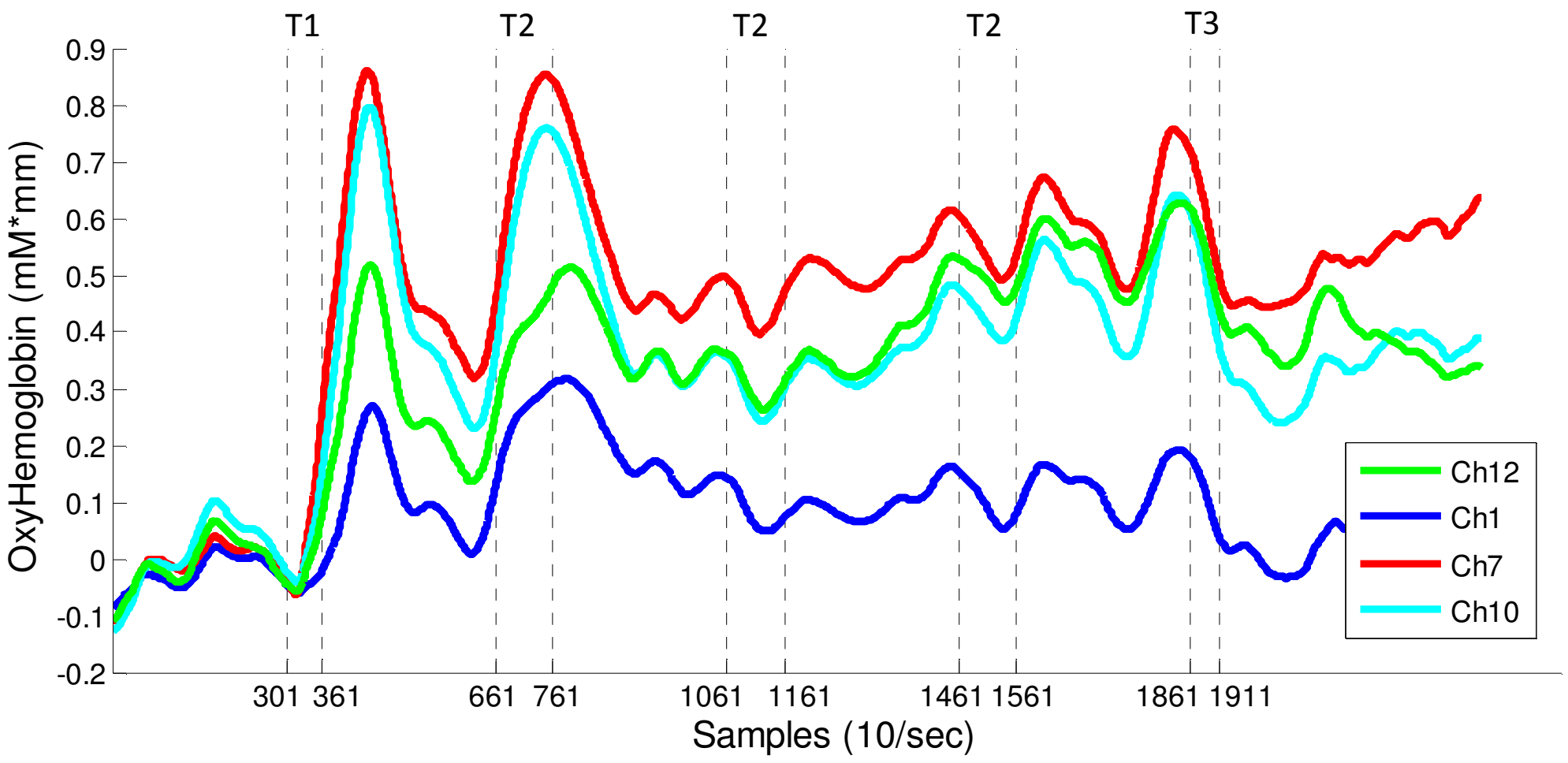

Figure 7. Pain Response after needle insertion.

Needle twirl (task 2) by the acupuncturist and it aims to increase the de-qi sensation. The time frame for this stimulation is ten seconds (samples 661-761) and was carried out immediately after resting time (Rt1), refer to Figure 6. The Acupuncturist stopped the twirl manipulation early when the patient expressed full numbness before the ten seconds. Similarly, the signal pain behaviour exhibited a small drop first before increasing constantly for 8 seconds. At approximately 14 seconds after the start of the twirl stimulation a large peak of HbO was obtained. Again, this signal dropped towards baseline after reaching its highest peak. After the stimulus application, the patient was given 30 seconds to rest.

The second (within samples 1061-1161) and third (within samples 1461-1561) twirl stimulation exhibited a different pattern than the first stimulation. These responses were smaller than the first response (see Figure 6), exhibiting a small increase and decrease of HbO after the stimulation. In contrast to the first needle twirl, the peak was reached after only three seconds in the second needle twirl; while for the last twirl, the stimulation response was slightly bigger than the previous twirl. The fact that the hemodynamic response decreased (compared to the first twirl stimulation) can be described by two factors: 1) the patient was familiar with the twirl manipulation and became adapted to this sensation, 2) by applying pressure in the hogu point (well known by its analgesic properties) the patient experienced a painkiller effect that decrease the effect of pain by needle manipulation. After the stimulus application, the patient was given 30 seconds to rest.

Hemodinamics of needle removal (task 3) are similar to task 1, in this stage the removal was done within a time frame of five seconds (samples 1861-1911). This stimulation was carried out immediately after the fourth resting time (Rt4). In this manipulation, the acupuncturist removed the needle with a fast abrupt pull. Similarly, to all previous stimulations, the pain response to needle manipulation exhibited a small drop of HbO followed by an increase to reach a peak in approximately four seconds after needle removal. After needle removal patient was given 40 seconds to rest which was used to let the signals go back to base line and to end the experiment.

### 3.3. Dominant pattern analysis

After visual analysis, lags between channels were observed which suggested dominant channels on both sides of the head. As shown in Figure 8, images taken after acupuncture stimulation show that around channel 7 (dotted lines) the magnitude of HbO is stronger and appears to be the centre of radiation towards surrounding channels. This is a strong indication that the area near channel 7 (Ch7) is a dominant area for this particular patient. Figure 8 also reveals that surrounding channels (Ch4, Ch5, Ch9 and Ch10) had a stronger relationship with Ch7 since their magnitude increased similarly. On the other hand, channels Ch1, Ch6 and Ch11 had a week relationship since their magnitude increased later at a slower rate.

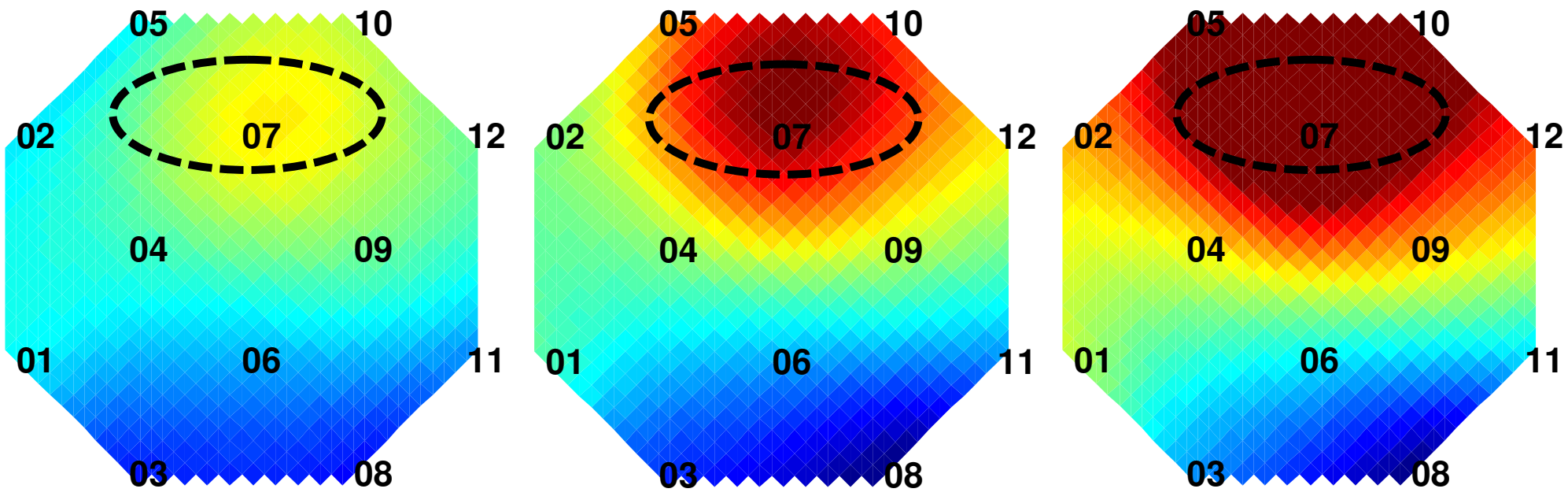

Figure 8. Images taken every five seconds after acupuncture stimulation showing dominant channel.

Based on the assumption that there is a relationship among channels, the cross correlation analysis was computed. Figure 9 shows the time-dependant cross correlation between Ch7 and other channels for needle insertion stimulus. The results show that channel 7 and surrounding channels have a small lag time that confirms the results of the visual analysis. An example of this is the lag between Ch7 and Ch9 ($\tau$=+0.2 sec), or Ch7 and Ch6 ($\tau$=+2.5 sec); these results reflect the lag time between this dominant region and the time delay among peripheral channels. In addition, no delay was observed between Ch7 and surrounding channels Ch5 and Ch10. These results suggest that the hemodynamic of stimulation signals have a progressive movement from dominant regions towards peripheral regions. It is also important to note that following stimulation tasks showed a similar pattern, however the lag time is less significant, it suggests that lag/led time between channels is directly related to the stimulation intensity.

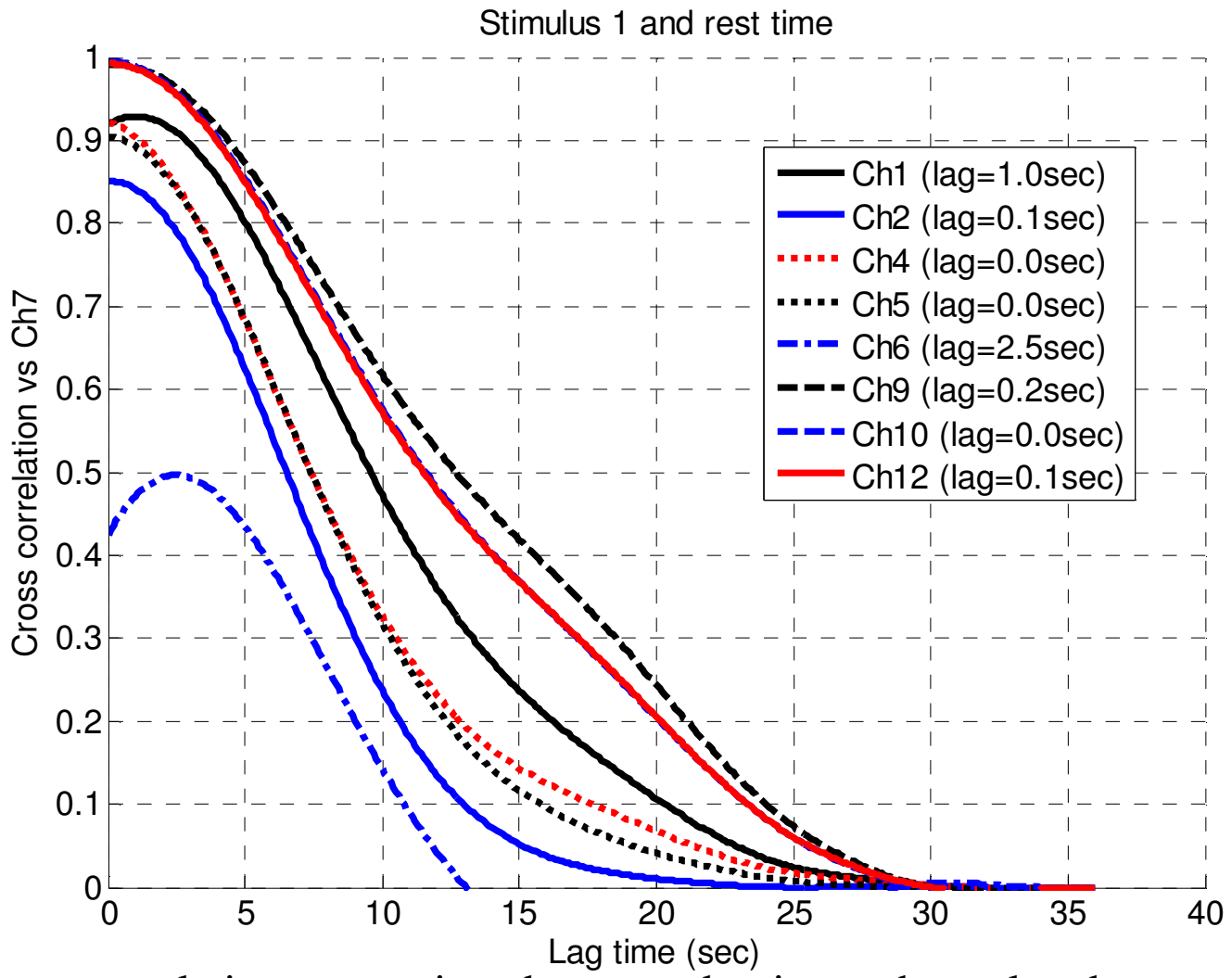

Figure 9. Cross correlation comparison between dominant channel and surrounding channels.

### 3.4. Direction of Pain Signal Information.

Following the results obtained by cross correlation analysis, optical flow was used to determine the direction of NIRS signals through the channels. Figure 10 shows the optical field computed with two consecutive images from channels 1-12 after needle insertion of patient four. It can be seen that flow of NIRS data goes from channel 7 and it radiates towards surrounding channels. It supports the previous finding using cross correlation analysis about the existence of dominant channels. This “dominant area” or “dominant channel” behaviour was found on both cerebral hemispheres and is present in all patients.

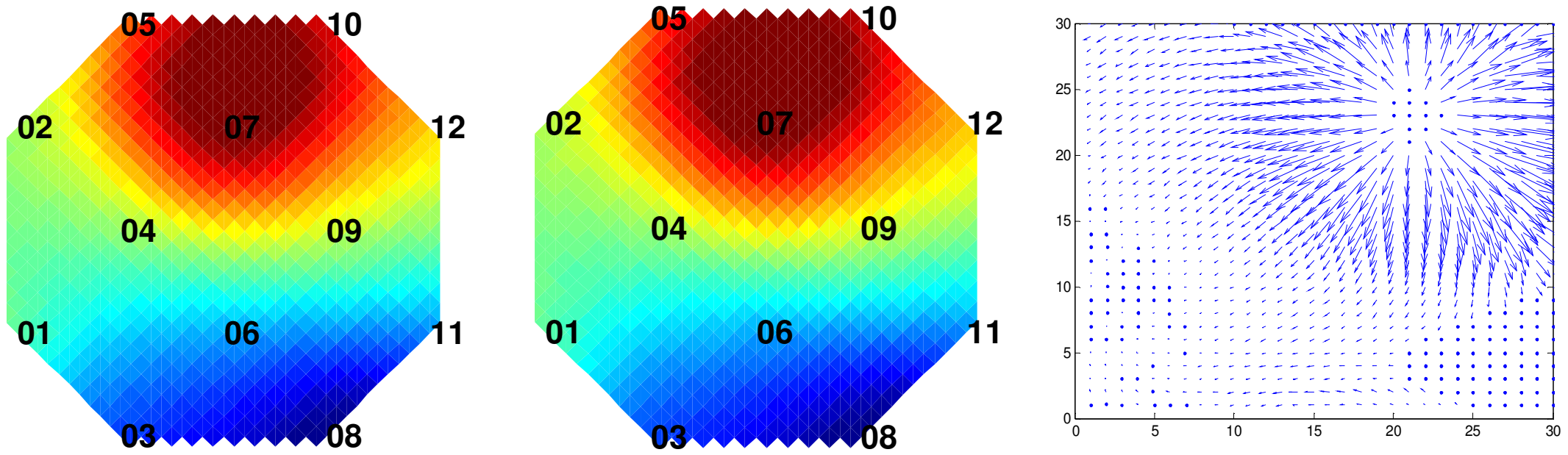

Figure 10. Optic flow computed from two consecutive images (samples 300 and 301) of patient four.

It is also important to note that based on optic flow analysis, pain relationship can be made. Figure 11 presents optic flow fields of four pair of samples taken from the complete acupuncture manipulation process. In this image, only four channels (Ch2, Ch7, Ch10, and Ch12) are showed in the image for demonstration purposes. Figure 11A shows the optic flow at four different stages through the experiment, the first two OF images exhibit the radiation pattern from Ch7 to peripheral channels; this outer movement is directly related to the slope increase of the pain data. On the other hand, the last two OF images represent an internal movement from peripheral channels to Ch7; this inner movement is related to the slope decrease in pain stimulation. In other words, it is possible to link the OF result with the intensity and progression of pain through the experiment.

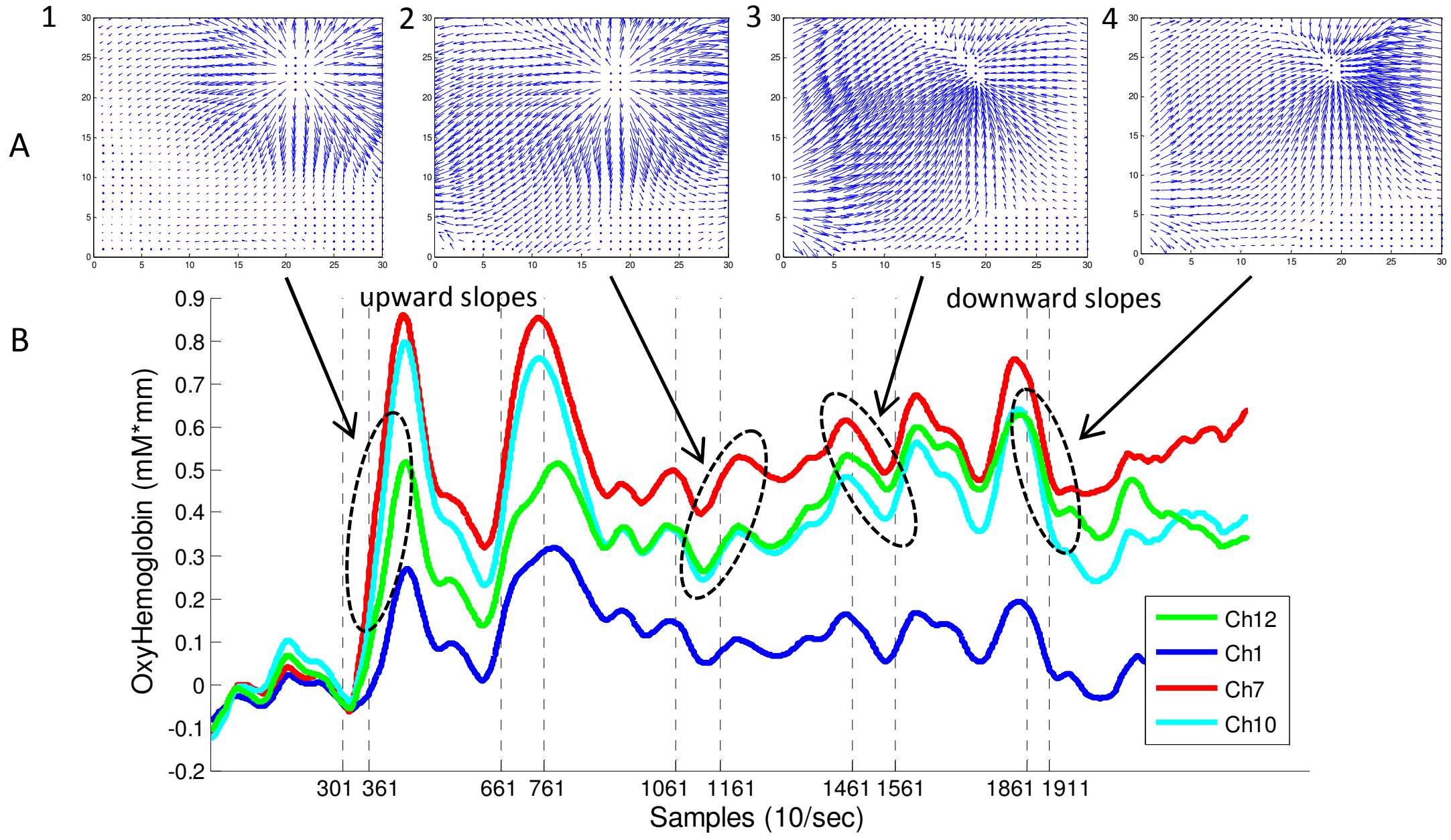

Figure 11. Optic flow results from four consecutive pair of samples, note the slope direction where the sample where taken. A) Optic flow fields showing the outer and inner flow of information between Ch7 and peripheral channels. B) NIRS data show the slope direction that reflects the intensity level of pain stimulation.

## 4. CONCLUSIONS

In this paper, the distinctive hemodynamic curve response to three types of acupuncture stimulation was presented. It was showed that the typical pain signal presents a small drop first to continue with a constant rise until it reaches a maximum to drop again towards resting level. This behaviour was consistent in all patients after the three types of acupuncture manipulations: needle insertion, needle twirl, needle removal.

We have also implemented cross correlation analysis and optical flow analysis within the context of diffuse optical imaging of acupuncture stimulation. Cross correlation confirmed visual analysis by identifying time-dependant relationship among channels. Regions where the neural activity is dominant, such as Ch7 in patient four, were also identified. This dominant neural activity was present in all patients, however magnitude levels varied among patients.

Optical flow corroborated pattern flow from dominant channels to peripheral channels. This method also distinguished the relationship between the pain stimulation and direction of pattern flow. It was found that an increase of pain stimulation was reflected as flow of NIRS information moving towards peripheral channels, while decrease of pain stimulation was reflected as flow of NIRS signals moving from peripheral channels to dominant channel.